\documentclass[aps,nofootinbib,superscriptaddress,twocolumn]{revtex4-1}
\usepackage{graphicx}
\usepackage{amsmath}
\usepackage{hyperref}
\usepackage{amssymb,latexsym,mathrsfs}

\hypersetup{
    colorlinks=true,
    linkcolor=red,
    citecolor=blue,
}

\def\be{\begin{equation}}
\def\ee{\end{equation}}
\def\ba{\begin{eqnarray}}
\def\ea{\end{eqnarray}}

\begin{document}


\title{Did BICEP2 see vector modes? First B-mode constraints on cosmic defects}

\author{Adam Moss}
\affiliation{School of Physics and Astronomy, University of Nottingham, University Park, Nottingham NG7 2RD, UK}
\author{Levon Pogosian}
\affiliation{Department of Physics, Simon Fraser University, Burnaby, BC, V5A 1S6, Canada}

\begin{abstract}
Scaling networks of cosmic defects, such as strings and textures, actively generate scalar, vector and tensor metric perturbations throughout the history of the universe. In particular, {\em vector} modes sourced by defects are an efficient source of the CMB B-mode polarization. We use the recently released BICEP2 and POLARBEAR B-mode polarization spectra to constrain properties of a wide range of different types of cosmic strings networks.  We find that in order for strings to provide a satisfactory fit on their own, the effective inter-string distance needs to be extremely large --  spectra that fit the data best are more representative of global strings and textures. When a local string contribution is considered together with the inflationary B-mode spectrum, the fit is improved. We discuss implications of these results for theories that predict cosmic defects.
\end{abstract}

\maketitle

Observations of the Cosmic Microwave Background (CMB) radiation have established a compelling case for an inflationary beginning of our universe \cite{Guth:1980zm,Linde:1981mu,Albrecht:1982wi,Starobinsky:1980te,Linde:1983gd,Vilenkin:1983xq}. Inflation resolved the monopole, the flatness and the horizon problems and, as a bonus, provided a mechanism for generating small fluctuations in the metric of space-time \cite{Starobinsky:1979ty,Mukhanov:1981xt,Hawking:1982cz,Starobinsky:1982ee,Bardeen:1983qw,Guth:1985ya,Mukhanov:1985rz}. CMB temperature anisotropies measured by COBE, WMAP and Planck are in a spectacular agreement with predictions of simplest inflationary models \cite{Bennett:1996ce,Bennett:2012zja,Ade:2013uln}. The same models also predict a scale-invariant spectrum of gravitational waves \cite{Starobinsky:1979ty} that can be imprinted in the CMB temperature and polarization \cite{Crittenden:1993wm,Frewin:1993dq,Harari:1993nb}. A smoking gun of the gravity waves is the so-called B-mode pattern of polarization \cite{Kamionkowski:1996zd,Seljak:1996gy,Zaldarriaga:1996xe,Kamionkowski:1996ks}. Recent data from BICEP2 \cite{Ade:2014xna} provided tantalizing evidence for this signal at an amplitude that is consistent with predictions of the simplest inflationary models. The B-mode signal seen by BICEP2 can also contain contributions from other sources. The purpose of this paper is to examine the implications of the BICEP2, as well as the recently released POLARBEAR results \cite{Ade:2014afa}, for another potential source of B-modes -- cosmic strings.

Particle theory suggests that the universe went through a series of symmetry breaking phase transitions as it expanded and cooled. Cosmic defects, such as monopoles, strings, domain walls and textures, could form in these phase transitions and potentially survive until the time of last scattering and even today \cite{Hindmarsh:1994re,Vilenkin:book}. Cosmic strings were actively studied as an alternative to inflation mechanism for generating the structure in the universe \cite{Vilenkin:1981iu}. Eventually, it became apparent that CMB and matter power  spectra predicted by cosmic strings were distinctly different from what was observed and they have been ruled out as the main seed for structure formation \cite{Albrecht:1997nt}. However, in models with multiple scalar fields, strings can form at the end of inflation \cite{Kofman:1995fi,Tkachev:1998dc} and contribute a small amount of power to the CMB temperature anisotropy \cite{bouchet02,Wyman:2005tu,Bevis:2007gh,Battye:2010hg,Battye:2010xz}. Such scenarios include supersymmetric grand unified models \cite{Jeannerot:1995yn,Jeannerot:2003qv,Rocher:2004my} and brane inflation \cite{Jones:2002cv,Sarangi:2002yt,Kachru:2003sx,PogosianTye,Dvali:2003zj}.

Among the predicted signatures of cosmic defects is the CMB B-mode polarization on sub-degree angular scales~\cite{Seljak:1997ii,Hu:1997hp,Battye:1998js,SelSlo,Bevis:2007qz,Pogosian:2007gi,Urrestilla:2007sf,Urrestilla:2008jv,Mukherjee:2010ve,Avgoustidis:2011ax}. The nature of cosmological perturbations generated by defects is qualitatively different from those set by inflation. The latter sets the initial conditions for the metric and matter inhomogeneities which subsequently evolve unperturbed. Defects, on the other hand, actively generate scalar, vector and tensor perturbations throughout the history of the universe \cite{Hu:1997hp,Pen:1997ae,Turok:1997gj}. Because vector modes quickly decay when not actively sourced, they are completely negligible in the inflationary mechanism. But, for defects, they are comparable to scalar modes and can be an efficient source of the CMB B-mode polarization. Tensor modes, or gravity waves, are also produced by defects but with a lower impact on CMB because of their oscillatory nature \cite{Hu:1997hp}.

In order for a string network to maintain scaling, long strings must chop off loops that subsequently radiate away. Pulsar timing measurements \cite{Jenet} and gravitational wave detectors \cite{Abbott:2009ws} strongly constrain the amount of gravitation waves produced by loops of {\em local} cosmic strings \cite{Damour:2001bk,Damour:2004kw,Siemens:2006vk,Siemens:2006yp}, giving bounds much tighter than the current CMB constraints. However, the amount of the gravitational wave emission from string loops and kinks is not as established \cite{Battye:2010xz,O'Callaghan:2010ww} as effects of the large scale dynamics of the string network on CMB. Also, the tight gravity wave bounds do not apply to {\em global} strings\footnote{We thank Alex Vilenkin for pointing this out.}, i.~e. those formed as a result spontaneously broken global (as opposed to local) gauge symmetries. We note that future B-mode experiments can come close to providing bounds \cite{SelSlo,Ma:2010yb,Avgoustidis:2011ax} comparable to those from gravity wave probes.

In this paper we use the newly released BICEP2 and POLARBEAR data to constrain properties of a wide variety of cosmic string networks. We consider two cases: one in which there is no contribution to B-modes from the inflationary gravity waves and one in which there is a mixture of the inflationary and string contributions. We provide quantitative answers to the following questions: 1) {\em Can cosmic strings provide a good fit to the BICEP2 and POLARBEAR B-mode spectra without any contribution from inflationary tensor modes?} 2) {\em Is the fit improved by adding a cosmic string contribution to the inflationary B-modes?} 3) {\em What are the implications of the new B-mode data for the properties of cosmic string networks?} 

In this paper we will refer to the tensor-to-scalar ratio $r$ evaluated at the scale $k=0.002\, {\rm Mpc}^{-1}$. To model the strings, we use the unconnected segment model (USM) \cite{Vincent:1996qr,ABR99,Pogosian:1999np,cmbact,Avgoustidis:2012gb}, which offers the ability to mimic the CMB spectra from different types of strings. The USM model was introduced in \cite{Albrecht:1997nt,ABR99}, based on the approach suggested in \cite{Vincent:1996qr}, developed into its present form in \cite{Pogosian:1999np}, and implemented in a publicly available code CMBACT \cite{cmbact}. The string unequal time correlators of the USM model can be derived using analytical expressions developed in \cite{Avgoustidis:2012gb}, which we use in this work.

In the USM, in addition to the dimensionless string tension $G\mu$, there are two important parameters -- the scaling parameter $\xi$, which sets the effective inter-string distance\footnote{$\xi \equiv La/\eta$, where $L$ is the mean inter-string distance (related to the string energy density via $\rho_s=\mu/L^2$), and $\eta$ is the conformal time. We stress that $\xi$ is an effective parameter in the USM model, and  $\xi>1$ does not necessarily imply presence of super-horizon correlations in the model whose spectra are reproduced by the USM.}, and the root-mean-square (RMS) velocity $v$. On cosmological scales, probed by the CMB measurements, the fine details of the string evolution do not play a major role. It is the large-scale properties, such as the scaling distance and the rms velocity, that determine the shape of the string-induced spectra. The overall normalization of the spectrum depends on G$\mu$ as well as the string number density, controlled by $\xi$.

The advantage of working with the USM is that one can quickly scan over spectra of many different types of cosmic defects to see if any of them happen to be favoured by data. Of course, this requires that USM is able to provide a satisfactory fit to the CMB spectra or, equivalently, to the stress-energy unequal time correlators (UETC), derived from available numerical simulations. For instance, it was shown in \cite{Battye:2010xz} that the USM can reproduce the CMB spectra derived from the simulations of local strings by \cite{Bevis:2007qz,Urrestilla:2007sf,Urrestilla:2008jv}. Fits to UETC from simulations by other groups \cite{Allen:1990tv,Fraisse:2007nu,BlancoPillado:2011dq} can also be performed, but are not available at this time. 

\begin{figure}[tb]
\includegraphics[width=1\columnwidth]{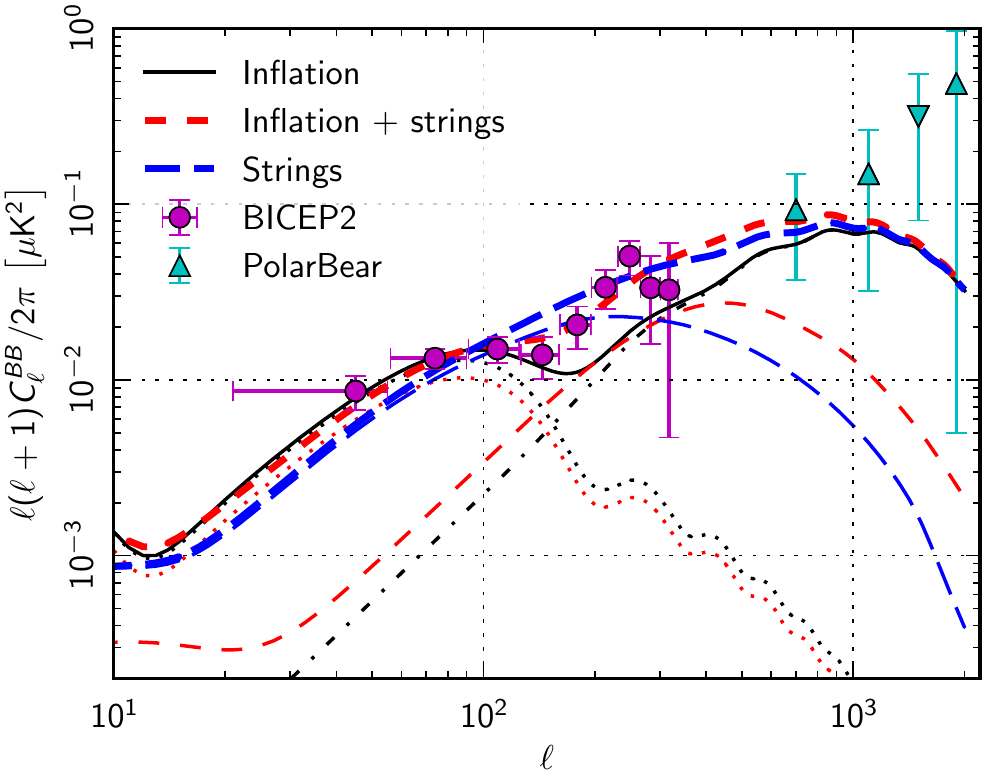}
\caption{The thick blue long dashed line is the best fit lensing+strings model ($r$=0), with the thin blue long dashed line showing the corresponding string contribution alone. The thick red short dash is the best fit lensing+strings+inflation model ($r$=0.15), with the corresponding string contribution plotted as a thin red short dashed line. The lensing contribution is shown separately with a thin black dot-dashed line. The BICEP2 best fit inflationary model ($r$=0.2) contribution is shown with a thin black dotted line, and the solid thin black line is the sum of $r$=0.2 and lensing contributions. The circles show the band powers measured by BICEP2 and the triangles are the POLARBEAR data (the third band is negative with its absolute value plotted as an inverted triangle).}
\label{fig:BB}
\end{figure}

The thin red short dashed line in Fig.~\ref{fig:BB} shows a typical B-mode spectrum generated by local strings. It is primarily sourced by {\em vector} modes and has two peaks. The less prominent peak at $\ell \sim 10$ is due to rescattering of photons during reionization, while the main peak, at higher $\ell$, is the contribution from last scattering. Both peaks are quite broad because a string network seeds fluctuations over a wide range of scales at any given time. The position of the main peak is determined by the most dominant Fourier mode stimulated at last scattering, which is set by the values of $\xi$ and $v$ \cite{Pogosian:2007gi}. The power tends to move to lower multipoles (larger angular scales) when either $v$ or $\xi$ are increased. Increasing $v$ also increases the width of the peak. In fact, because $v<1$ sets a maximum scale, it takes a large increase in $\xi$ to move the peak to the left (to lower $\ell$) even by a small amount.

Let us briefly comment on how we quantify the string contribution to CMB. Bounds on cosmic string are often quoted solely in terms of $G\mu$. Such bounds implicitly assume the scaling configuration of local strings in the Abelian Higgs model, where at any time there is roughly one Hubble length string per Hubble volume. More generally, the bound on strings depends on the combination of  $G\mu$ and the string number density\footnote{In the one-scale model, $N_s \propto \xi^{-3}$. However, the inter-string distance, which can be very small in models with lower intercommutation probabilities, need not to be the same as the coherence scale along the string, which remains of ${\cal O}(H^{-1})$.} $N_s \propto \xi^{-2}$. Typically,  $\xi \lesssim 1$, but can be much smaller in models with lower intercommuting probabilities. Moreover, different types of observations probe different combinations of $\xi$ and $\mu$. As shown in \cite{PogosianTye}, CMB power spectra (and other two-point correlation functions) constrain $\mu \sqrt{N_s} \sim \mu/\xi$, while gravity wave probes essentially constrain the string energy density given by $\mu/\xi^2$. To avoid the model-dependence when interpreting the CMB bounds in terms of $G\mu$, we follow the Planck Collaboration  \cite{Ade:2013xla} and quantify the amount of the anisotropy contributed by strings in terms of $f_{10}$, which is the fractional contribution of strings to the CMB temperature spectrum at $\ell=10$, $f_{10} \equiv C^{\rm str}_{10} / C^{\rm tot}_{10}$. The first year Planck data constrains it at $f_{10} \lesssim 0.03$ \cite{Ade:2013xla}. In this work we do not fit to Planck data, instead focusing on the implications of the B-mode data alone. Values of $f_{10}$ that exceed Planck bounds can be disregarded. We also ignore the small theoretical uncertainty involved in calculating the lensing contribution to the B-mode spectrum.

\begin{figure}[tb]
\includegraphics[width=0.9\columnwidth]{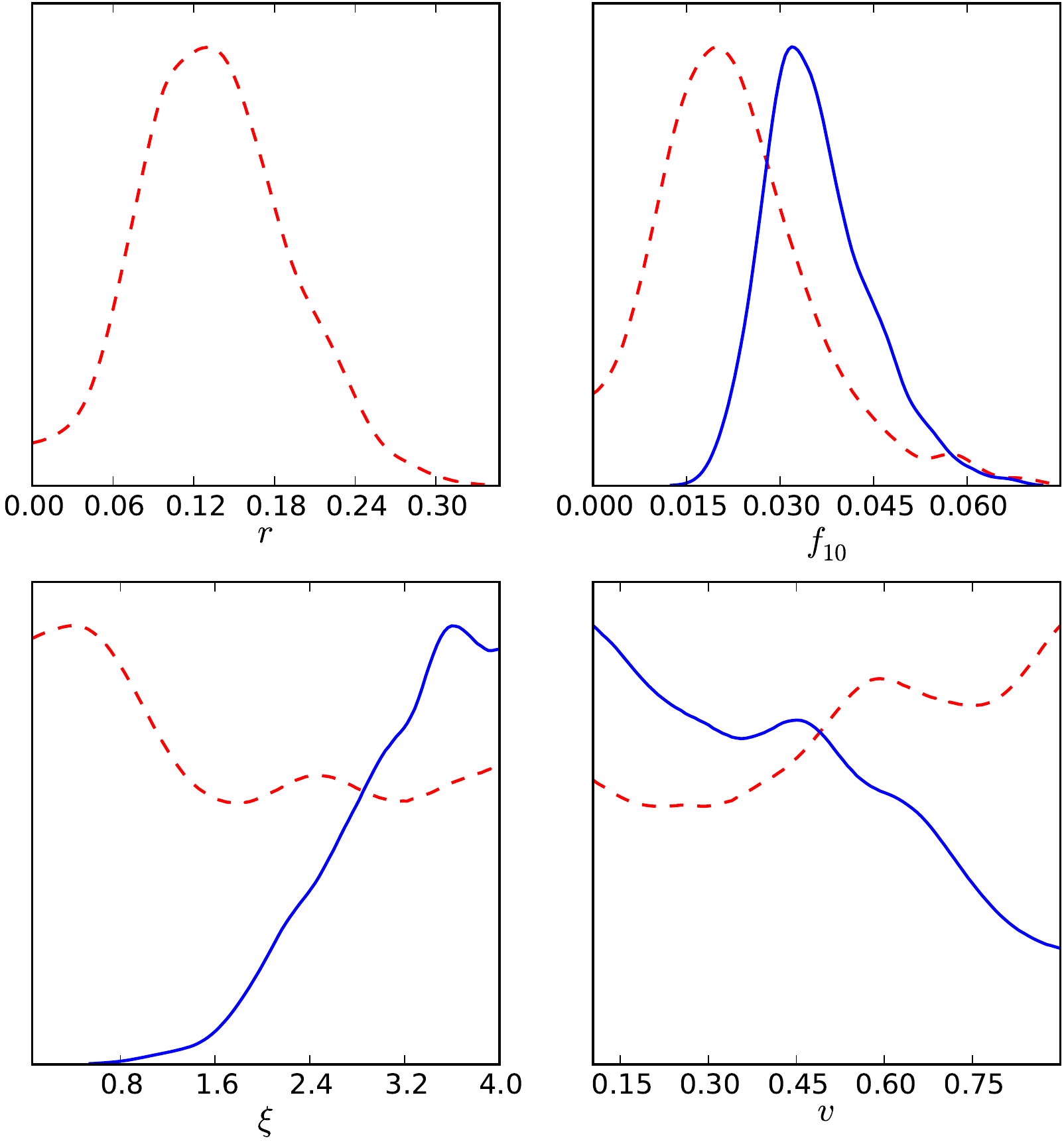}
\caption{Marginalized likelihoods derived from the BICEP2 and POLARBEAR data for the scalar-to-tensor ratio $r$, the strength of the string contribution $f_{10}$, the inter-string distance $\xi$, and the RMS velocity $v$. The red dotted lines are for the lensing+strings+inflation model, while the blue solid lines are for the lensing+strings fit only.}
\label{fig:1D}
\end{figure}

We first discuss how the string {\em only} model compares to inflation. The blue solid lines in Fig.~\ref{fig:1D} show the marginalized likelihoods of $f_{10}$, $\xi$ and $v$ obtained by fitting the string B-mode spectra, combined with lensing, to the BICEP2 and POLARBEAR data. There is a well-defined peak at $f_{10}=0.036 \pm 0.008$, with preference for larger $\xi$ values. The overall $\chi^2$  is only slightly worse ($\Delta \chi^2 = 2.65$) compared to inflation, although the string model has 2 additional parameters. The corresponding string contribution to the B-mode spectrum is shown with a thin blue long dash line in Fig.~\ref{fig:BB}, and strings+lensing with a thick blue long dash line. Not surprisingly, the data, which has  a bump at $\ell \sim 100$, favours a spectrum with a peak at a lower $\ell$, which is at $\ell \sim 250$ for the best fit model. A model with such large values of $\xi$ corresponds to rare and heavy strings -- the implied value of $G\mu$ in this model is $5\times 10^{-6}$, but their number density is low, which allows it to remain consistent with Planck bounds. The peak position for this model is closer to that of global strings and textures \cite{Pen:1997ae,Urrestilla:2007sf}, and certainly not representative of local strings \cite{Battye:2010xz}. This is also clear from the likelihood plot for $\xi$, which effectively {\em rules out} models with $\xi < 1.8$ $(2 \sigma)$ as the only primordial source of B-modes. For reference, the B-mode spectra from the local string simulation of \cite{Bevis:2007qz} correspond to the USM with $\xi \approx 0.4$ \cite{Battye:2010xz}. We can foresee that models with global strings, textures \cite{Pen:1997ae,Urrestilla:2007sf} or global phase transitions, of the kind discussed in \cite{JonesSmith:2007ne,Dent:2014rga}, would provide a much better fit than local strings.

We now consider the model in which both strings {\em and} inflation generate B-modes. The red dotted lines in Fig.~\ref{fig:1D} show the marginalized likelihoods of $r$ and the string parameters in a model with an additional inflationary tensor mode contribution. In this case, the fit is improved relative to the model with no strings, with $\Delta \chi^2 =-6.06$ and 3 additional parameters. The marginalized string fraction is $f_{10}=0.025 \pm 0.014$, which corresponds to $G\mu \approx 4 \times 10^{-7}$, with slight preference for lower values of $\xi$, characteristic for local strings, and $r= 0.14 \pm 0.05$. 

\begin{figure}[tbh]
\includegraphics[width=0.9\columnwidth]{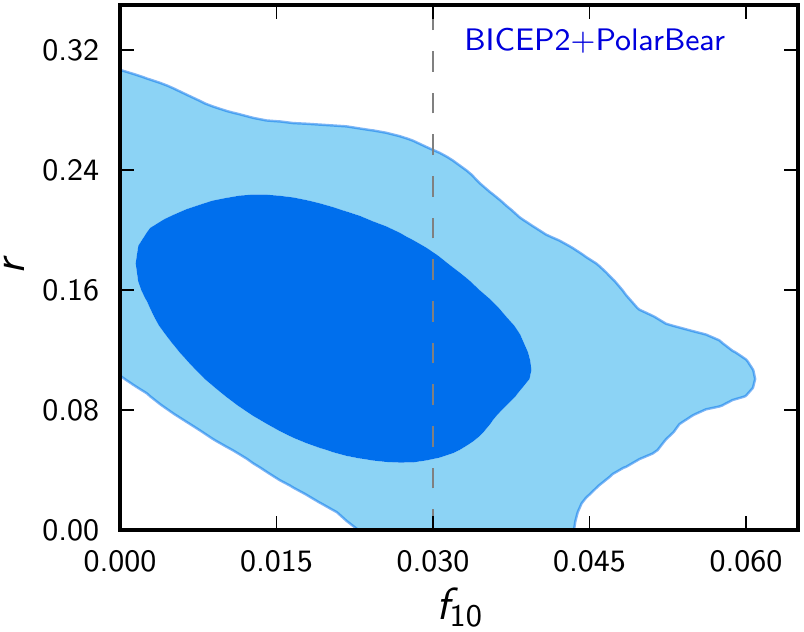}
\caption{The marginalized joint likelihood for the tensor-to-scalar ratio $r$ and the strength of the string contribution $f_{10}$. The two different shades indicate the $68$\% and the $95$\% confidence regions. The vertical dashed line indicates the approximate bound on $f_{10}$ from Planck.}
\label{fig:2D}
\end{figure}

Fig.~\ref{fig:2D} shows the marginalized joint likelihood for $r$ and the strength of the string contribution $f_{10}$. It clearly shows that a combination of the two contributions fits the data better than when either of them is zero. The thin vertical line indicates the approximate upper bound on $f_{10}$ from Planck\footnote{The Planck Collaboration did not scan over all values of $\xi$ and $v$, instead it provided two separate bounds on $f_{10}$ corresponding to two different string models. We quote the weaker of the two bounds because both models were included in the USM parameter space covered by our fit.}. It should be noted that the improvement in the fit comes primarily from data points at higher $\ell$, while the BICEP2 collaboration warns \cite{Ade:2014xna} that points at $\ell>150$ should be considered as preliminary.

Our findings carry implications for models that predict defects. The inability of {\em local} strings to fit the B-mode spectrum {\em on their own} poses a problem for the simplest and most studied brane inflation models in which inflation ends with a production of cosmic superstrings~\cite{Sarangi:2002yt,PogosianTye,Kachru:2003sx}. Such models predict tiny values of $r$, and the only observable B-modes could come from strings, which are effectively of local, Nabmu-Goto, type. The fact that local strings do not fit the BICEP2 data puts these scenarios under pressure. Generally, since $r$ tends to be small in hybrid inflation-type models, it is not clear if inflationary models with such large values of $r$ can be consistent with production of cosmic strings. 

The main reason cosmic strings struggle to provide a good fit to the BICEP2 data is the presence of B-mode power on smaller angular scales. This power could be suppressed if strings were to form not after but during inflation~\cite{Lazarides:1984pq,Shafi:1984tt}. Such strings could remain far-separated and prevented from reaching a scaling solution until the onset of decoupling\footnote{This idea was pointed out to us by Alex Vilenkin}. A related scenario has recently been discussed in \cite{Kamada:2012ag} as a way of eliminating the presence of loops during the radiation era and, thus, evading the tight pulsar bounds on cosmic strings. There may be an impetus for investigating such models further in the context of string-sourced B-modes.

To summarize, we have shown that the B-mode spectra measured by BICEP2 and POLARBEAR are consistent with a contribution from vector modes sourced by cosmic strings. Working with the USM model allowed us to scan over a wide range of scaling defect models parametrized by the effective density parameter $\xi$ and the RMS velocity $v$. In order for strings to provide a satisfactory fit to data on their own, the $\xi$ parameter needs to be extremely large, well beyond values typical for local strings. The string spectra that fit the data best are more representative of global strings and textures.

When the string contribution is considered together with the inflationary B-mode spectrum, they improve the overall fit. This is primarily because the string contribution allows the model to pass through the data points at $\ell>150$. The best fit USM model in this case is consistent with B-mode spectra from simulations of local strings.

In both cases, with and without the inflationary contribution, the best fit for $f_{10}$ is close to, but still below the bound set by Planck based on fits to the CMB temperature spectra. Thus, we expect that a joint fit that included the Planck data would not significantly change the conclusions of this paper. Such a fit must be performed in the future when more data becomes available.

We have argued that detectable B-modes can be produced by cosmic defects. Other interesting possibilities include phase transitions \cite{Dent:2014rga} and primordial magnetic fields \cite{Mack:2001gc,Lewis:2004ef}. Thus, BICEP2 results are exciting not only because of the potential discovery of the signal from inflationary gravity waves, but also because they have pioneered the era of precision B-mode science -- a new frontier for testing fundamental physics with cosmology.

{\it Acknowledgements}~~We acknowledge helpful discussions with Richard Battye, Ed Copeland, Antony Lewis, Carlos Martins, Dani Steer, Henry Tye, Tanmay Vachaspati, Alex Vilenkin, Ira Wasserman and Mark Wyman. We specially thank Antony Lewis for help with the likelihood code. LP is supported by an NSERC Discover Grant. AM is supported by STFC. 

{\em While this paper was in preparation, a related short paper was posted on arxiv.org \cite{Lizarraga:2014eaa} commenting on similar ideas. Our work provides quantitative answers to some of the questions posed in \cite{Lizarraga:2014eaa}.}

\end{document}